\documentclass[twocolumn,showpacs,preprintnumbers,amsmath,amssymb]{revtex4}
\usepackage{graphicx}
\usepackage{dcolumn}
\usepackage{bm}
\usepackage{morefloats}
\usepackage{color}

\begin{document}

\preprint{APS PRA}

\title{The photon scattering cross-sections of atomic hydrogen}
\author{Swaantje~J.~Grunefeld}
\affiliation{School of Mathematics and Physics, The University of Queensland, Brisbane, St Lucia, QLD 4072, Australia}
\author{Yongjun~Cheng}
\affiliation{Academy of Fundamental and Interdisciplinary Science, Harbin Institute of Technology, Harbin 150080, PR China}
\affiliation{School of Engineering, Charles Darwin University, Darwin NT 0909, Australia}
\author{Michael~W.~J.~Bromley}
 \homepage{www.smp.uq.edu.au/people/brom/}
\affiliation{School of Mathematics and Physics, The University of Queensland, Brisbane, St Lucia, QLD 4072, Australia}
\date{\today}

\begin{abstract}

We present a unified view of the frequency dependence of the various
scattering processes involved when a neutral hydrogen atom interacts with a
monochromatic, linearly-polarized photon.
A computational approach is employed of the atom trapped by a
finite-sized-box due to a finite basis-set expansion, which generates a set of
transition matrix elements between $E<0$ eigenstates and $E>0$ pseudostates.
We introduce a general computational methodology that enables the
computation of the frequency-dependent dipole transition polarizability
with one real and two different imaginary contributions.
These dipole transition polarizabilities are related to the
cross-sections of one-photon photoionization, Rayleigh, Raman,
and Compton scattering.  Our numerical calculations reveal individual
Raman scattering cross-sections above threshold that can rapidly vanish and revive.
Furthermore, our numerical Compton cross-sections do not overtly
suffer from the infra-red divergence problem, and are
three orders-of-magnitude higher than previous analytic-based Compton scattering
cross-sections.  Our total photon-hydrogen scattering cross-sections 
thus resolve the discrepancies between these previous calculations
and those in the N.I.S.T. `FFAST' database.

\end{abstract}


\maketitle


The problem of photon-atom scattering was first tackled using
quantum theory in the mid-to-late 1920s, resulting in the
development of the Kramers-Heisenberg-Waller matrix
elements~\cite{sakurai67a}.  These describe the fundamental
Rayleigh and Raman processes, however, are also known to
suffer from an infra-red divergence problem when computing
Compton scattering cross-sections~\cite{heitler54a}.
In this paper we avoid this problem by building a
computational photon-plus-atom-in-a-box, which effectively
results in an upper photon wavelength based on the size of
the box, and we are able to obtain the total
photon-hydrogen scattering cross-sections.

The real and two different imaginary dipole transition polarizabilities
are set up in the present methodology and used to compute the set of 
non-relativistic low-\textit{frequency} 
photon-hydrogen cross-sections.
The incident photon field \textit{strengths} are assumed to lie in the
weak-to-intermediate regime where a collision involving two-incident photons
are unlikely, and where second-order perturbative treatments of a photon-atom
collision are applicable~\cite{delone00a}.
Photon-atomic hydrogen experiments are notoriously challenging~\cite{fried98a}.
Previously three-photon ionization experiments have been performed~\cite{kryala91a},
and recently experiments have been performed further into the strong-field
regime using ultrafast laser pulses~\cite{kielpinski14a}, neither of which do
we attempt to connect our results to since (high-order perturbative) multiphoton
treatments are required.

\begin{figure}[bth]
\caption{Schematic of some photon scattering processes from the hydrogen $1s$ state.
The incoming photon frequency, $\omega$, is depicted here as lying
above the ionization threshold.
Series (a) indicates Rayleigh scattering where one of the perturbative
terms involves the physical ($4p$) eigenstate with the absorption / spontaneous
emission of an $\omega$ photon. 
The allowed dipole decays from the $4p$ state are shown in dashed-lines.
Series (b) shows Compton scattering, where one of the perturbative
terms involves an $E>0$ $\ell=1$ pseudostate, and then spontaneous decay
of frequency $\omega^\prime$ shown down into a $\ell=2$ pseudostate
(which would subsequently ionize). 
Some allowed dipole decays from the $\ell=1$ pseudostate
 are shown in dashed-lines. These impact the linewidths of these
processes (during Compton scattering the final $\ell=2$ pseudostate would
also have a number of decay channels that impacts its linewidth).}
\label{fig:schematic}
\vspace{0.1cm}
\includegraphics[width=85mm,angle=0]{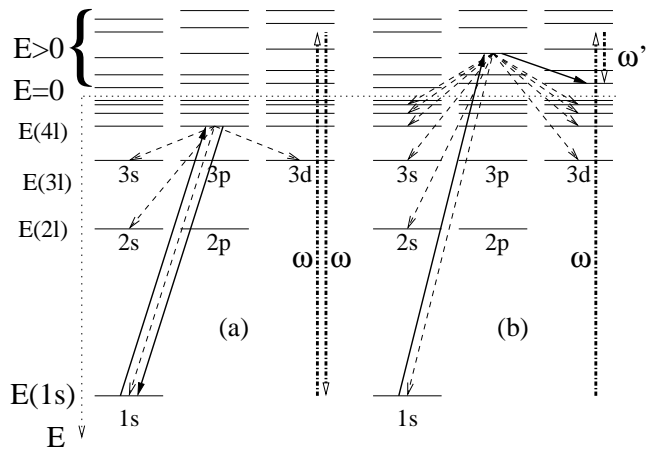}
\end{figure}

The fundamental computational approach taken in this paper to compute
the cross-sections is to use the transition matrix elements connecting
two states via a complete set of intermediate states summed over both
bound eigenstates and pseudostates.
These are schematically shown in Fig.~\ref{fig:schematic} for the cases of Rayleigh and
Compton scattering where an $\ell = 1$ 
 state is given as an example of the
intermediate state and some possible dipole decay pathways from this state
are also shown that would impact the transition linewidth.
We present cross-sections here without resolving either fine or
hyperfine structure, which in future work could be included~\cite{delserieys08a}.
%

This paper was initially motivated by the incompleteness in the
compiled set of theoretical photon-hydrogen total cross-sections
in Fig.~29 of Bergstrom \textit{et al.}~\cite{bergstrom93a}.
There they did not present Raman scattering cross-sections,
whilst Compton cross-sections were presented over a limited frequency
range based on a low-energy (infra-red) photon truncation of the analytic
differential cross-sections from Gavrila~\cite{gavrila69b,gavrila72a,gavrila74a}.
They consequently noted that ``the total Compton cross-section
does not continue to fall at low energies''~\cite{bergstrom93a} as one
would predict.

This problem was more recently tackled
by Drukarev \textit{et al.}~\cite{drukarev10a},
who also used the work of Gavrila, and also an approximation
to numerically avoid the infra-red divergence.
Their total Compton cross-section does fall at low energies,
however, their highest-energy cross-sections at $100$~eV
are mismatched by an order-of-magnitude against those
in Ref.~\cite{bergstrom93a}.
Both of these calculations, furthermore,
lie vastly below those in the
``FFAST: Form Factor, Attenuation, \& Scattering Tables"
computed by Chantler~\cite{chantler95a}.
The FFAST database covers all atoms and are used for a variety of applications
spanning medical and materials science applications~\cite{pratt04a,pratt14a}.
Our results in this paper are qualitatively
similar to, yet $30-60$\% larger than,
the FFAST scattering cross-sections, and this is just
for the neutral hydrogen atom.  

\textit{Photon-Atom Methodology} ---
Our numerical method begins by diagonalizing our hydrogen
atom using a finite-sized (orthogonal) Laguerre basis set that
provides some kind of a soft-walled potential atom-in-a-box~\cite{mitroy08e}.
This discretizes the continuum and provides both bound state and
`pseudostate' information that can be used
to compute freqency-dependent polarizabilities
below the ionization threshold~\cite{mitroy10a,tang13a},
and for dispersion coefficients~\cite{jiang15a}.
The pseudostate information can also be exploited to
perform, eg. lepton-atom scattering~\cite{mitroy07a,mitroy08e}.
The use of pseudostates for above threshold photon-atom
scattering was explored by Langhoff \textit{et al.}
for Rayleigh scattering and one-photon
photoionization~\cite{langhoff70a,langhoff73a,langhoff74a,langhoff74b,langhoff76a}.

We extend these methods to also perform calculations of
Raman scattering and, furthermore, of Compton scattering.
The photon-atom scattering cross-sections are based on the
Kramers-Heisenberg-Waller matrix elements involving
the electromagnetic coupling
$H_c = (2mc^2)^{-1}e^2A^2 - (mc)^{-1}e \vec{p}\cdot\vec{A}$~\cite{heitler54a,sakurai67a}.
We ignore the $A^2$ `seagull' term (the Waller matrix element) since that is
only important at $\gtrsim$~keV photon energies~\cite{eisenberger70a,bergstrom93a}.
The Kramers-Heisenberg matrix element is determined here
as a transition polarizability, $\alpha_{ji}(\omega)$, between some
initial state $|i;L_iS\rangle$ and final state $|j;L_jS\rangle$ through
a complete set of intermediate states $|t;L_tS\rangle$~\cite{chandrasekharan81a}.
The details of our algorithms are given in the Supplemental Material, Sections I and II.
In brief, we use reduced matrix elements,
assuming linear polarization~\cite{bonin97a,delserieys08a}, such that
\begin{equation}
\begin{split}
  \alpha_{ji}(\omega) \approx \sum_{t}  C_{L_i,L_t,L_j} \Bigg[ & \frac{\langle j || z || t \rangle \langle t || z ||i \rangle}
                                            {\varepsilon_{ti}-\omega-i \frac12 \Gamma_{ti}(\omega)} \\
              +                           & \frac{\langle j || z || t \rangle \langle t || z || i \rangle}
                                            {\varepsilon_{tj}-(-\omega)-i \frac12 \Gamma_{tj}(\omega)} \Bigg] ,
  \label{eqn:alphatrans}
\end{split}
\end{equation}
where $\varepsilon_{ab} = E_b - E_a$, and all quantities above
are in atomic units (a.u.).  We use the expressions in Ref.~\cite{delserieys08a},
to find that $C_{0,1,0} = \frac13$, whilst $C_{0,1,2} = \frac13$.

Each $\alpha_{ji}(\omega)$ is calculated as a sum over intermediate states $t$, 
which are either bound or pseudostates. When the intermediate state is a bound 
state, we compute an imaginary term (denoted $\mathrm{Im}_0$) by the damping of 
the oscillator through the linewidth/s of the
atomic bound states~\cite{heitler54a,bonin97a,lekien13a,lepers14a}. 
Wijers has argued~\cite{wijers04a} that the decay rates
must be frequency dependent to ensure
$\Gamma_{ab}(\omega) \to 0$ as $\omega \to 0$.
Thus we use 
%
%
$\Gamma_{ab}(\omega) = (\Gamma_a + \Gamma_b)
                     (2\varepsilon_{ab}^2 \omega^2/(\varepsilon^4_{ab} + \omega^4))$.
This form of the resonant damping includes the case where
the state that absorbs the photon has a non-zero decay rate to
other bound states.

If the intermediate state in Eqn.~\ref{eqn:alphatrans} is
a pseudostate, then the linewidth is not included 
as continuum states do not have a physical linewidth.
For $\omega > |E_{1s}|$, this results in unphysical 
singularities in the continuum when $\omega = \varepsilon_{ti}$.
These are removed by assuming infinitesimally small pseudostate linewidths $\Gamma_{ti}\to 0^+$,
which results in real and imaginary (denoted $\mathrm{Im}_1$) terms~\cite{langhoff74a}.
This enables us to compute one real and the two different imaginary polarizabilities.
Our calculations of $\alpha_{ji}$ where $i \equiv 1s$
were all performed with a fixed number of Laguerre-type orbitals $N_\ell=120$ for each angular momentum,
which gives $18$ $\ell=1$ $(E<0)$ bound states and $102$ $\ell=1$ $(E>0)$ pseudostates.
Convergence studies against other basis sets are relegated to Supplemental Section IV.

\begin{figure}[tbh]
\caption{Rayleigh and photoionization scattering cross-sections for
photon-hydrogen scattering (in units of $\sigma_T$).
The $\sigma_{e}$ and $\sigma_{I}$ are shown as a function of
incident photon energy $\omega$ in atomic units
(ie. up to $\approx 27$~eV).  Our $\sigma_{e}$ results agree with the
analytic results of Gavrila~\cite{gavrila67b}, with some discrepancy
against the (as digitized by us) numerical results of Bergstrom \textit{et al.}~\cite{bergstrom93a}.
The FFAST $\sigma_I$ results lie a few percent below ours~\cite{chantler95a}.
}
\includegraphics[width=85mm]{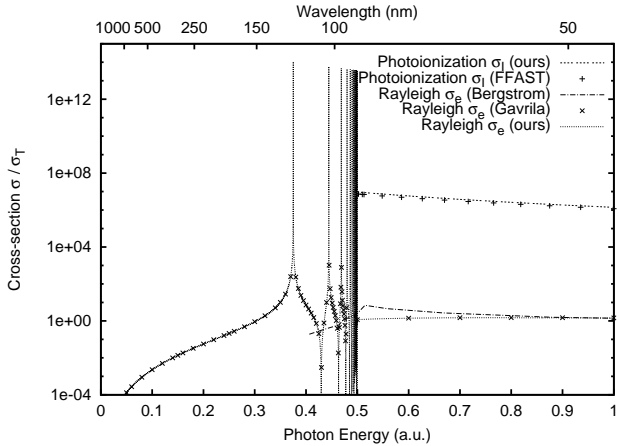}
\label{fig:rayleigh}
\end{figure}
\textit{Cross-section results} ---
For (elastic) Rayleigh scattering of a photon with frequency $\omega$ we have~\cite{langhoff74a,sadeghpour92b,delserieys08a}
\begin{equation}
   \sigma_e(\omega) = \sigma_T \omega^4 \Big|\mathrm{Re}\left[\alpha_{ii}(\omega)\right]
                                          + i\mathrm{Im}_0\left[\alpha_{ii}(\omega)\right]\Big|^2 ,
  \label{eqn:crossrayleigh}
\end{equation}
where $\sigma_T$ is the Thomson scattering cross-section
of a photon with a free electron
(for reference, $\sigma_T \approx 6.65 \times 10^{-25}$ cm$^2$),
whilst $\omega$ and $\alpha_{ii}$ are both in a.u..
The photoionization cross-section is given by~\cite{langhoff74b}
\begin{equation}
  \label{eqn:crossoptical1}
   \sigma_{I}(\omega) = \sigma_T \frac32 c^3 \omega \: \mathrm{Im}_1\left[\alpha_{ii}(\omega)\right] ,
\end{equation}
i.e. the optical theorem. The speed-of-light, $c\approx 137$ (in a.u.),
gives the massive enhancement factor of $\sigma_{I}$ over that of $\sigma_e$.
Our results for these cross-sections for validation purposes are shown in Fig.~\ref{fig:rayleigh},
where our results agree with the (not-shown) analytic $\sigma_I$ function~\cite{sobelman96a}.
Fig.~\ref{fig:rayleigh} also shows physical resonances and
that $\sigma_e(\omega\to\infty) \approx \sigma_T$ in the non-relativistic limit. 
The Rayleigh scattering cross-sections between $\omega = 0.37-0.55$~a.u.,
and up to $\omega = 100$~a.u., are given in Supplemental Figs.~1 and 3.

\begin{figure}[tbh]
\caption{Raman scattering cross-sections for
photon-hydrogen scattering (in units of $\sigma_T$).
These are shown as a function of incident photon energy $\omega$ in
atomic units (ie. up to $\approx 27$ eV).
Our results are compared against the available ($\omega < 0.5$)
results for H($1s$) $\to$ H($2s$) of Sadeghpour and Dalgarno~\cite{sadeghpour92b}.
The cross-section for H($1s$) $\to$ H($3d$) is also shown.
The sum of the individual cross-sections $\sigma_{R}(\omega)$ is also shown
as the solid line, and is seen to monotonically decrease.
}
\includegraphics[width=85mm]{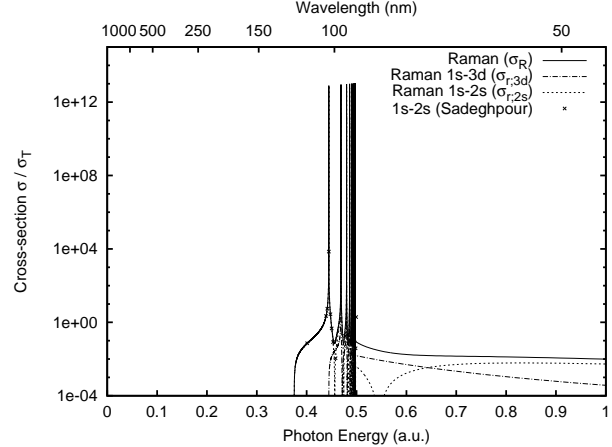}
\label{fig:raman}
\end{figure}

For (inelastic) Raman scattering to (physical) state $j$
\begin{equation}
   \sigma_{r;j}(\omega) = \sigma_T \; \omega \left(\omega_j^\prime \right)^3
                 \Big|\mathrm{Re}[\alpha_{ij}(\omega)] + i \mathrm{Im}_0[\alpha_{ij}(\omega)]\Big|^2 ,
  \label{eqn:crossrayram}
\end{equation}
with spontaneously emitted photon frequency $\omega_j^\prime = \omega - \omega_{ij}$.
Thus the total Raman cross-section for an atom in an initial state $i$ is
$\sigma_{R}(\omega) = \sum_{(j;E_j < \Delta)} \sigma_{r;j}(\omega)$
(see supplemental material for our basis-set-based choice of
ionization location $E = \Delta$, which effectively demarcates the
bound states from the pseudostates).  Our Raman codes were
able to be validated at energies below ionisation $\omega < 0.5$~a.u.,
where previous photon-hydrogen Raman calculations have computed
H($1s$) $\to$ H($2s$) scattering~\cite{sadeghpour92b},
and excited initial state H($3s$) $\to$ H($3d$) `Rayleigh' scattering~\cite{florescu85a}.
Our H($1s$) $\to$ H($2s$), H($1s$) $\to$ H($3d$),
and total Raman cross-sections are shown in Fig.~\ref{fig:raman}.  
The individual H($1s$) $\to$ H($ns$) Raman cross-sections above
threshold rapidly vanish and revive as each of their matrix elements
pass through a different `tune-out' wavelength where $|\alpha_{ij}(\omega)|\approx 0$.
The H($1s$) $\to$ H($nd$) cross-sections are monotonically
decreasing above threshold.
The total Raman cross-section above $\omega = 0.5$~a.u.
monotonically decreases such that $\sigma_{R}(\omega) \to 0$ as $\omega \to \infty$
(see Supplemental Figs.~2 and 4 for cross-sections for $\omega = 0.37-0.55$~a.u.,
and up to $\omega = 100$~a.u.).
%
%

The final process considered here is that of
Compton scattering, which opens up for frequencies above threshold $\omega>|E_i|$,
and analytically requires the \textit{differential} cross-section
to be integrated over all possible outgoing photon frequencies $\omega^\prime$~\cite{drukarev10a},
\begin{equation}
   \sigma_{C}(\omega) = \int_{\omega^\prime_{\mathrm{min}}}^{\omega^\prime_{\mathrm{max}}} \frac{d \sigma_C}{d\omega}\Bigg|_{\omega^\prime} d\omega^\prime
                         \approx \sum_{(j;E_j>\Delta)} \sigma_{r;j}(\omega) ,
   \label{eqn:completecompton}
\end{equation}
where the largest emitted photon frequency $\omega^\prime_{\mathrm{max}} = \omega-|E_i|$,
whilst the smallest $\omega^\prime_{\mathrm{min}} = 0$.  The infra-red problem is
that $\frac{d \sigma_C}{d\omega}$ diverges as $\omega^\prime \to 0$, resulting
in an infinite cross-section, and thus previous analytic calculations
of Bergstrom~\textit{et al.}~\cite{bergstrom93a} assumed $\omega^\prime_{\mathrm{min}} = 10$~eV,
whilst Drukarev~\textit{et al.}~\cite{drukarev10a} assumed $\omega^\prime_{\mathrm{min}} = 1$~eV.
We, instead, adopt the approximation in Eqn.~\ref{eqn:completecompton}, adapting the same
formulae from the Raman case since the finite set of pseudostates gives us a discrete sum. 
Note that our Raman vs Compton delineation is in the same spirit of previous work
on excitation Raman vs ionization Raman photon-helium scattering~\cite{grosges99a}.
The convergence of the sum towards the integral can be understood by considering that,
as the basis size $N_\ell$ is increased, more pseudostates are included
and the magnitude of the individual cross-sections $\sigma_{r;j}(\omega)$ decreases,
whilst the $\sigma_{C}(\omega)$ tends to remain constant (see Supplemental Fig.~5).

\begin{figure}[tbh]
\caption{Comparisons of Compton scattering cross-sections up
to high energies for photon-hydrogen scattering (in units of $\sigma_T$).
The summed $\sigma_{C}(\omega)$ is shown as a function of
incident photon energy $\omega$ in atomic units
(ie. up to $\approx 2700$ eV).
The summed cross-section is also shown when only the
$L_j = 0$ contributions are included, which approximately
agrees with the results of Bergstrom~\textit{et al.}~\cite{bergstrom93a} and
Drukarev~\textit{et al.}~\cite{drukarev10a}
(their results were digitized by us).
The FFAST results are also shown~\cite{chantler95a}.
}
\includegraphics[width=85mm]{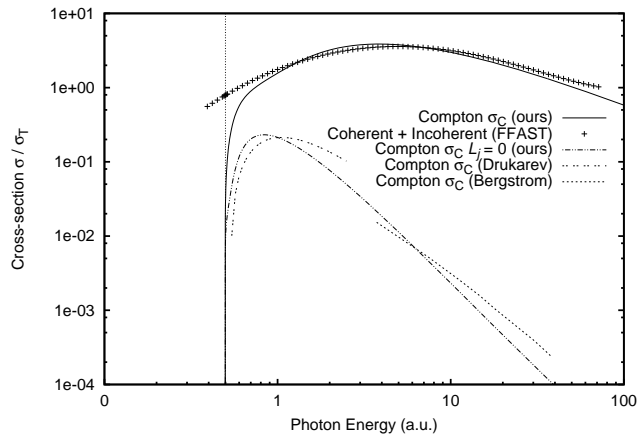}
\label{fig:comptoncompare}
\end{figure}
Our results for Compton scattering are shown in Fig.~\ref{fig:comptoncompare}
where our results completely disagree by over three orders-of-magnitude with
the previous results of Bergstrom~\textit{et al.}~\cite{bergstrom93a}
and Drukarev~\textit{et al.}~\cite{drukarev10a}.  We, however, ran
our calculations by including only the states with final $L_j = 0$,
ie. ignoring the $L_j = 2$ contributions which turn out to dominate
the sum in Eqn.~\ref{eqn:completecompton}.
In doing so we find qualitative agreement of our $0\to 1\to 0$ cross-section
with these previous calculations, which relied on numerical truncation
of analytic differential cross-sections.
Our total results instead broadly agree with the FFAST
coherent + incoherent results~\cite{chantler95a}.
Coherent scattering generally refers to elastic (Rayleigh) scattering,
whilst incoherent scattering means inelastic (Compton) scattering~\cite{pratt14a}.

\begin{figure}[t]
\caption{Comparisons of scattering cross-sections up
to high energies for photon-hydrogen scattering (in units of $\sigma_T$).
These are shown as a function of incident photon energy
$\omega$ in atomic units (ie. up to $\approx 2700$ eV).
The coherent, incoherent, and photoionization data from FFAST~\cite{chantler95a}
and XCOM~\cite{berger10a} is shown, as well as photoionization from
XAAMDI~\cite{hubbell04a}.  Note that the $y$-axis is here shown not on logscale.
}
\includegraphics[width=85mm]{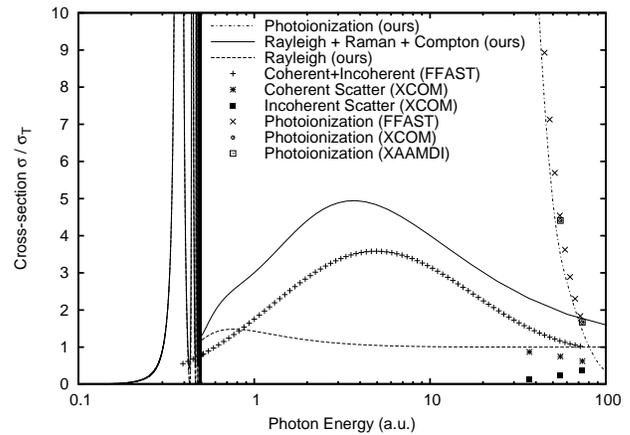}
\label{fig:total}
\end{figure}

\textit{Total Cross-section} ---
The total cross-section summing all Rayleigh, Raman, Compton processes
is shown in Fig.~\ref{fig:total}, showing qualitative agreement with
the FFAST data. The NIST-based XCOM and XAAMDI database results are
presented in Fig.~\ref{fig:total}~\cite{hubbell75a,hubbell79a,berger10a,hubbell04a}.
The photoionization cross-section only appears in Fig.~\ref{fig:total} 
towards keV energies where it drops down in magnitude to be below the others.
Our data shows that Wentzel's rule~\cite{kaplan76a} is never applicable
to hydrogen, that is, the sum over the elastic
and inelastic total cross-sections does not tend to that of a free electron
(ie. the Thomson cross-section) over the energy range before higher-order
effects take over~\cite{bergstrom93a}.

\textit{Conclusion} --- We have introduced a computational method
for computing various cross-sections of photon-atom scattering
for any initial state from a single atomic structure calculation.
We find the expected behaviour of the Compton scattering
towards low-energies and our calculations do not appear to suffer from
the infra-red catastrophe.  Effectively a computational atom-in-box
sets a maximum wavelength that can be `measured'~\cite{gavrila72a,gavrila72b}.
Our results are able to reach up to energies where the
beyond-dipole-approximation and relativistic/retardation effects
become important and thus provide a benchmark for future
work.  Our methods can be extended to compute the cross-sections
for atoms in various initial states, and where knowledge of the
possible scattering processes will help guide experimentalists when
designing atomic and molecular experiments.
In particular, it will be worthwhile to compute Rayleigh, Raman, and
Compton scattering cross-sections for multi-electron atoms where
Cooper minima in the photoionization cross-sections and resonances
can occur~\cite{pratt14a}.


\begin{acknowledgments}

The work of MWJB was supported by an Australian Research Council
Future Fellowship (FT100100905).
YC was supported by the National Natural
Science Foundation of China (Grant No. 11304063).
The work of YC was also partially enabled by a Discovery Project (DP-1092620)
of Prof. Jim Mitroy (deceased).
We thank Dr. Julian Berengut, A/Prof. Tom Stace,
 and Prof. Dave Kielpinski for useful conversations and 
Dr. Sergey Novikov for correspondence about matrix elements.

\end{acknowledgments}


\end{document}


\preprint{APS PRA}

\title{Supplemental data --- The photon scattering cross-sections of atomic hydrogen}
\author{Swaantje~J.~Grunefeld}
\affiliation{School of Mathematics and Physics, The University of Queensland, Brisbane, St Lucia, QLD 4072, Australia}
\author{Yongjun~Cheng}
\affiliation{Academy of Fundamental and Interdisciplinary Science, Harbin Institute of Technology, Harbin 150080, PR China}
\affiliation{School of Engineering, Charles Darwin University, Darwin NT 0909, Australia}
\author{Michael~W.~J.~Bromley}
 \email{brom@physics.uq.edu.au}
\affiliation{School of Mathematics and Physics, The University of Queensland, Brisbane, St Lucia, QLD 4072, Australia}
\date{\today}

\begin{abstract}

We present: in section (I) some formulae underlying our computations,
wherein we extend the treatment by Delserieys \textit{et al.}~\cite{delserieys08a}
to handle the above threshold region,
in section (II) discussion of how we determine which states are bound
and which are treated as pseudostates, and finally
in section (III) some zoomed in plots spanning the physical resonance area,
and in section (IV) some demonstrations of the convergence of our
photon-hydrogen scattering cross-sections with respect to the
Laguerre basis employed here.

\end{abstract}


\maketitle

\section{Formulae}

Our atom-in-a-box calculation generates a large number of transition matrix elements.
These connect the $N_\ell$ number of Laguerre basis functions for each
partial wave to each other through dipole allowed reduced matrix elements
of operator $\hat{z} \equiv r C^1(\hat{r})$,
that are independent of the magnetic quantum number. 
Given $C^1(\hat{r})$ is the spherical tensor of rank-$1$,
for eigenstate $i$ and eigenstate $j$, 
the reduced matrix elements are
\begin{equation}
  \begin{split}
  \langle j|| r C^1(\hat{r}) || i \rangle
 &=   \int\int Y_{\ell_j}(\theta,\phi) C^1(\theta,\phi) Y_{\ell_i}(\theta,\phi) \sin \theta d\theta d\phi
     \int \psi_j(r) r \psi_i(r) r^2 dr \\
 &=   (-1)^{\ell_j} \sqrt{(2\ell_j+1)(2\ell_i+1)}
      \left( \begin{array}{ccc}
        \ell_j & 1 & \ell_i \\
           0   & 0 & 0
      \end{array} \right)
     \int \psi_j(r) r \psi_i(r) r^2 dr  \; .
  \label{eqn:reduced}
  \end{split}
\end{equation}
%

The orthogonal Laguerre basis functions that we choose to use all
have the same long-range parameterisation, $\exp(-\lambda_\ell r$),
for each partial-wave, $\ell$.  By providing a set of basis functions with fixed $\lambda = 0.5$, 
they are explicitly not a set of eigenstates of hydrogen 
(as hydrogen eigenstates have $\lambda_n = \frac12 n$~\cite{bromley03a}).
Thus we generate a set of
 low-lying bound states as well as a
(reasonably well-defined) set of higher-lying Rydberg states,
along with a set of ($E>0$) pseudostates that reflect the finite
nature of the effective box induced by having a finite number of
basis functions.

In our code the radial integrals are performed numerically~\cite{bromley01a},
which has allowed us in the past to perform numerous calculations
of one- and two-electron atoms including mixtures of Laguerre and Slater
Type Orbitals to represent frozen Hartree-Fock-based core electrons~\cite{bromley01a,mitroy03e}.
Here we use a large radius grid spanning to maximum of $\approx 1000$ a.u.
using $4096$ grid points with $16$-point Gauss-Legendre radial integration quadrature~\cite{bromley01a}.

The `Kramers-Heisenberg' matrix element is determined here
as a transition polarizability between some initial state
$|i;L_iS\rangle$ and final state $|j;L_jS\rangle$ through a complete
set of intermediate states $|t;L_tS\rangle$.
In terms of sums over reduced matrix elements with
$LS$-coupled wavefunctions and assuming linear photon polarization~\cite{delserieys08a},
\begin{equation}
\begin{split}
  \alpha_{ji}(\omega) & = \sum_{t}^{\infty}  C_{L_i,L_t,L_j} \Bigg[ \frac{\langle j || z || t \rangle \langle t || z ||i \rangle}
                                            {\varepsilon_{ti}-\omega-i \frac12 \Gamma_{ti}(\omega)} \Bigg]
                        + \sum_{t}^{\infty}  C_{L_i,L_t,L_j} \Bigg[ \frac{\langle j || z || t \rangle \langle t || z || i \rangle}
                                            {\varepsilon_{tj}-(-\omega)-i \frac12 \Gamma_{tj}(\omega)} \Bigg] \\
                      & + \int_{0}^{\infty}  \frac{C_{L_i,L_\epsilon,L_j}}{\rho(\epsilon)}
                                            \Bigg[  \frac{\langle j || z || \epsilon \rangle \langle \epsilon || z ||i \rangle}
                                            {\varepsilon_{\epsilon i}} \Bigg] d\epsilon
                       + \int_{0}^{\infty}  \frac{C_{L_i,L_\epsilon,L_j}}{\rho(\epsilon)}
                                            \Bigg[  \frac{\langle j || z || \epsilon \rangle \langle \epsilon || z || i \rangle}
                                            {\varepsilon_{\epsilon j}-(-\omega)} \Bigg] d\epsilon ,
  \label{eqn:alphatrans}
\end{split}
\end{equation}
where, in atomic units, $\varepsilon_{ab} = E_b - E_a$.
The bound eigenstates, denoted by $t$, form an infinite set. 
As the sums have bound intermediate states with physical linewidths, the decay rates $\Gamma_{ti}$ or $\Gamma_{tj}$ are included. 
The integrals describe the transitions where the intermediate states are continuum states, denoted by energy $\epsilon$. 
The transition matrix elements in the continuum integrals are normalized by the energy density $\rho(\epsilon)$ 
of the continuum states.
The terms with $-\omega$ in the denominator indicate photon absorption followed by emission.
The terms with $-(-\omega)$ in the denominator indicate photon emission followed by absorption.
For the H($1s$) initial state considered here,
$C_{0,1,0} = \frac13$ and also $C_{0,1,2} = \frac13$, where
we only have intermediate $L_\epsilon = L_t = 1$.
Whilst Eqn.~(\ref{eqn:alphatrans}) is for the Raman case, it covers
the Rayleigh case where $j = i$.

Here we break up the terms into the bound and continuum contributions. We sum up the bound states using 
the notation $T_{ba} = \langle b || z || a \rangle$, as
 \begin{equation}
        \mathrm{Re}_0\left(\alpha_{ji}(\omega)\right) + i \mathrm{Im}_0\left(\alpha_{ji}(\omega)\right) 
        \approx \sum_{b=1}^{N_b} C_{L_i,L_b,L_j} \Bigg[
                                 \frac{T_{jb} T_{bi}}{\varepsilon_{bi}-\omega-i \frac12 \Gamma_{bi}(\omega)}
                               + \frac{T_{jb} T_{bi}}{\varepsilon_{bj}-(-\omega)-i \frac12 \Gamma_{bj}(\omega)} \Bigg],
 \end{equation}
%
where $N_b$ is the finite number of (intermediate) bound states in a given calculation. 

Next we consider the first continuum integral in Eqn.~(\ref{eqn:alphatrans}), both below and above ionization threshold.
For the first continuum integral in Eqn.~(\ref{eqn:alphatrans}), when $\omega$ lies below the ionization threshold,
\begin{equation}
       \mathrm{Re}_1\left(\alpha_{ji}(\omega)\right)
       = \int_{0}^{\infty}  \frac{C_{L_i,L_\epsilon,L_j}}{\rho(\epsilon)}
             \Bigg[ \frac{T_{j\epsilon} T_{\epsilon i}}{\varepsilon_{\epsilon i}-\omega} \Bigg] d\epsilon
       \approx \sum_{p=1}^{N_p} \frac{C_{L_i,L_p,L_j}}{\Delta \varepsilon_p}
             \Bigg[ \frac{T_{j p} T_{p i}}{\varepsilon_{p i}-\omega} \Bigg] \Delta \varepsilon_p ,
\end{equation}
%
where $\rho(\epsilon) \approx \Delta \varepsilon_p$. Note that the sum over the pseudostates
is only an approximation to the integral, however, 
the sum tends to monotonically converge as the basis set increases.
When $\omega$ lies above the ionization threshold, 
unphysical poles occur at $\omega = \varepsilon_{\epsilon i}$.
We have to integrate around the pole by
introducing a small complex term into the frequency $z = \omega + i0^+$ where the $+$ indicates
approaching zero from above.  This means that we apply the Cauchy principal theorem,
\begin{equation}
\begin{split}
       \mathrm{Re}_1\left(\alpha_{ji}(z)\right) - i \mathrm{Im}_1\left(\alpha_{ji}(z)\right)
        = \int_{0}^{\infty}  
              \frac{\mathcal{F}(\epsilon)}{\varepsilon_{\epsilon i}-\omega-i 0^{+}}  d\epsilon
        \equiv \mathcal{P} \! \! \! \int_{0}^{\infty} 
              \frac{\mathcal{F}(\epsilon)}{\varepsilon_{\epsilon i}-\omega}  d\epsilon
        - i  \pi \mathcal{F}(\epsilon)\bigg|_{\epsilon=\omega + E_i},
\end{split}
\end{equation}
where $\mathcal{F}(\epsilon) = C_{L_i,L_\epsilon,L_j} T_{j\epsilon} T_{\epsilon i} / \rho(\epsilon)$. 
Again, however, we have a discrete set of pseudostates. Thus we evaluate the above
threshold values \textbf{only} at discrete frequencies $\omega_{ni} = E_n - E_i$,
which correspond to the frequencies required to resonantly excite the pseudostates,
and we remove the single singular term from the sum:
\begin{equation}
       \mathrm{Re}_1\left(\alpha_{ji}(\omega_{ni})\right) - i \mathrm{Im}_1\left(\alpha_{ji}(\omega_{ni})\right)
        \approx 
        \left(\sum_{p \ne n}^{N_p} \frac{C_{L_i,L_p,L_j}}{\Delta \varepsilon_p}
             \Bigg[ \frac{T_{j p} T_{p i}}{\varepsilon_{pi}-\omega_{ni}} \Bigg]\Delta \varepsilon_p \right)
          - i \pi C_{L_i,L_n,L_j} \frac{T_{j n} T_{n i}}{\frac12\left(E_{(n+1)}-E_{(n-1)}\right)} ,
\end{equation}
with modification of the finite differencing when computing at the first $n=1$ or the last $n=N_p$ pseudostates.

The second continuum integral in Eqn.~(\ref{eqn:alphatrans}),
corresponds first to photon emission, followed by absorption and
thus has no such pole due to the continuum.  This integral can be
evaluated both above and below threshold at any incident photon energy as
\begin{equation}
     \mathrm{Re}_2\left(\alpha_{ji}(\omega)\right) = \int_{0}^{\infty}  \frac{C_{L_i,L_\epsilon,L_j}}{\rho(\epsilon)}
                                            \Bigg[  \frac{\langle j || z || \epsilon \rangle \langle \epsilon || z || i \rangle}
                                            {\varepsilon_{\epsilon j}-(-\omega)} \Bigg] d\epsilon
                                               \approx \sum_{p=1}^{N_p} \frac{C_{L_i,L_p,L_j}}{\Delta \varepsilon_p}
                                              \Bigg[ \frac{T_{j p} T_{p i}}{\omega_{pj}+\omega} \Bigg]\Delta \varepsilon_p .
\end{equation}
%
This term only contains a divergence
for the case of resonant `anti-Stokes' Raman processes, where the final state lies at
an energy below the initial state, which are not considered here.

Thus for either Rayleigh ($i=j$) or Raman ($i \ne j,$ where $j \in [1,N_b]$) we can compute the
transition polarizability at any frequency $\omega$ below threshold:
\begin{equation}
   \alpha_{ji}(\omega) = \mathrm{Re}_0\left(\alpha_{ji}(\omega)\right) + \mathrm{Re}_1\left(\alpha_{ji}(\omega)\right) + \mathrm{Re}_2\left(\alpha_{ji}(\omega)\right)
                       + i \mathrm{Im}_0\left(\alpha_{ji}(\omega)\right),
\end{equation}
whilst for $\omega$ above threshold, we evaluate at the discrete pseudostate frequencies
\begin{equation}
   \alpha_{ji}(\omega_{ni}) = \left[\mathrm{Re}_0\left(\alpha_{ji}(\omega_{ni})\right) + \mathrm{Re}_1\left(\alpha_{ji}(\omega_{ni})\right) + \mathrm{Re}_2\left(\alpha_{ji}(\omega_{ni})\right)\right] + i \mathrm{Im}_0\left(\alpha_{ji}(\omega_{ni})\right) - i \mathrm{Im}_1\left(\alpha_{ji}(\omega_{ni})\right).
\end{equation}

For the case of Compton scattering, which is only available when the final state lies
in the continuum (ie. $j \to \epsilon_j$),
it can only be computed when the incident photon frequency is $\omega \geq \varepsilon_{ji}$.
However, we simply apply all of the same pseudostate machinery as outlined above when computing these
transition polarizabilities. That is, we compute the integrals by assuming that both the
intermediate states and final states are discrete pseudostates
and evaluate these
at the discretized frequencies $\omega_{ni} = E_n - E_i$,
where $n$ is the intermediate $L=1$ pseudostate.

\section{Choice of which eigenstates are bound vs pseudostates}

Since we are computing the non-relativistic hydrogen atom, with the
basis set employed here we obtain $E_{1s} = -0.5$~a.u. to near machine
precision.  The natural ionization potential is then located
exactly at $E = 0$.  Table~\ref{tab:transitiondata} shows the
transition data from the H($1s$) state to the lowest $25$ $p$-states
from the largest $N=120$ calculation.  This table shows that in
a single-shot Laguerre calculation we are able to accurately
reproduce up to approximately the $n=15$ Rydberg state.
Looking at the oscillator strength pattern, it reaches a minimum
for the $n=15$ state, after which the oscillator strengths start
increasing again. Not shown is that they reach another maximum
at the $85$-th $\ell=1$ eigenstate ($E_{86p} = 0.409641$~a.u.,
with $f_{if} = 0.007975$), before dropping down towards zero.
This corresponds to photon frequencies of $\omega \approx 0.9$~a.u.,
which is approximately where the (above threshold) Rayleigh and
Compton scattering peaks occur.
\begin{table}[h]
\caption{The $1s-np$ transition data for the first $25$ $\ell=1$
eigenstates from our $N=120$ Laguerre basis set calculation,
whose results are compared to the exact Rydberg values.
The $n_i \to n_f$ column indicates the initial to final state transition.
The $E_f$ (calc) column gives the eigenenergies in atomic units
of the final state from the calculation, whilst $E_f$ (exact) column gives
the exact energies based on the Rydberg formula ($E_{n} = -0.5/n^2$).
The transition oscillator strengths $f_{if}$ (calc) are from the
calculation, whilst the $f_{if}$ (exact) are from the
analytic formulae~\cite{bethe57a}.
The final column gives the relative difference between
the oscillator strengths
($=(f_{if} \mathrm{(calc)}-f_{if} \mathrm{(exact)})/f_{if} \mathrm{(exact)}$).
Two row demarcations are given, the first to indicate where the
$E<0$ bound states have a minima in the oscillator strengths and
start to lose accuracy, the second where the $E>0$ pseudostates begin.
}
\label{tab:transitiondata}
\vspace{0.2cm}
\begin{tabular}{l|c|c|c|c|c}
\hline \hline
$n_i \to n_f$ & $E_f$~(calc) & $E_{f}$~(exact) & $f_{if}$~(calc) & $f_{if}$~(exact) & (rel. diff.) \\
\hline
$1s \to 2p$  & -0.125000 & -0.125000 & 0.416197 & 0.416197 & 0.000000 \\
$1s \to 3p$  & -0.055556 & -0.055556 & 0.079102 & 0.079102 & 0.000000 \\
$1s \to 4p$  & -0.031250 & -0.031250 & 0.028991 & 0.028991 & 0.000000 \\
$1s \to 5p$  & -0.020000 & -0.020000 & 0.013938 & 0.013938 & 0.000000 \\
$1s \to 6p$  & -0.013889 & -0.013889 & 0.007799 & 0.007799 & 0.000000 \\
$1s \to 7p$  & -0.010204 & -0.010204 & 0.004814 & 0.004814 & 0.000000 \\
$1s \to 8p$  & -0.007813 & -0.007813 & 0.003183 & 0.003183 & 0.000000 \\
$1s \to 9p$  & -0.006173 & -0.006173 & 0.002216 & 0.002216 & 0.000000 \\
$1s \to 10p$ & -0.005000 & -0.005000 & 0.001605 & 0.001605 & 0.000000 \\
$1s \to 11p$ & -0.004132 & -0.004132 & 0.001201 & 0.001201 & 0.000000 \\
$1s \to 12p$ & -0.003472 & -0.003472 & 0.000921 & 0.000921 & 0.000000 \\
$1s \to 13p$ & -0.002959 & -0.002959 & 0.000723 & 0.000723 & 0.000076 \\
$1s \to 14p$ & -0.002550 & -0.002551 & 0.000582 & 0.000577 & 0.007629 \\
$1s \to 15p$ & -0.002206 & -0.002222 & 0.000531 & 0.000469 & 0.132738 \\
\hline
$1s \to 16p$ & -0.001850 & -0.001953 & 0.000609 & 0.000386 & 0.578355 \\
$1s \to 17p$ & -0.001428 & -0.001730 & 0.000723 & 0.000321 & 1.253279 \\
$1s \to 18p$ & -0.000933 & -0.001543 & 0.000831 & 0.000270 & 2.073712 \\
$1s \to 19p$ & -0.000372 & -0.001385 & 0.000930 & 0.000230 & 3.050423 \\
\hline
$1s \to 20p$ &  0.000254 & -         & 0.001024 &  - &  - \\
$1s \to 21p$ &  0.000940 & -         & 0.001116 &  - &  - \\
$1s \to 22p$ &  0.001688 & -         & 0.001205 &  - &  - \\
$1s \to 23p$ &  0.002496 & -         & 0.001293 &  - &  - \\
$1s \to 24p$ &  0.003365 & -         & 0.001381 &  - &  - \\
$1s \to 25p$ &  0.004295 & -         & 0.001468 &  - &  - \\
$1s \to 26p$ &  0.005287 & -         & 0.001555 &  - &  - \\
\hline
\hline
  \end{tabular}
\end{table}

We were able to use Table~\ref{tab:transitiondata} to demarcate where the
bound states end and where the pseudostates begin.  We choose this energy
to be just above where the oscillator strength reaches a minima.
For the present $N_\ell=120$ calculations this is at
$\Delta_{120} \approx (-0.002206+-0.001850)/2 = -0.002028$ a.u..
The Raman final states are thus those with $E_n < -0.00203$, whereas the
Compton have $E_n > -0.00203$.
This choice, rather than the exact $\Delta =0$, means that our effective
ionization potential in our box has a small residual error in it.
However, the moving of the states with $\Delta_{N_\ell} < E_n < 0$ to be
counted as Compton rather than Raman removes spurious cross-sections
that were polluting the convergence of the total Raman cross-section
at photon-frequencies above threshold.  Essentially, the wavefunctions
corresponding to the $16p$ to $19p$ states appear to contain a
significant mixture of both Rydberg and continuum information.
This does mean that the $Im_1$ contribution also starts at
similarly lower photon energy $0.5-0.00203=0.4979$ a.u.,
and it does mean that our calculations have a large error
in this relatively small energy range around threshold.

\section{A zoomed in view of the cross-sections}

\begin{figure}[tbh]
\caption{Rayleigh scattering cross-sections for
photon-hydrogen scattering (in units of $\sigma_T$).
The $\sigma_{e}$ are shown as a function of
incident photon energy $\omega$ in atomic units.  Our $\sigma_{e}$ results agree with the
analytic results of Gavrila~\cite{gavrila67b}, with some discrepancy
against the (as digitized by us) numerical results of Bergstrom \textit{et al.}~\cite{bergstrom93a}.}
\includegraphics[width=85mm]{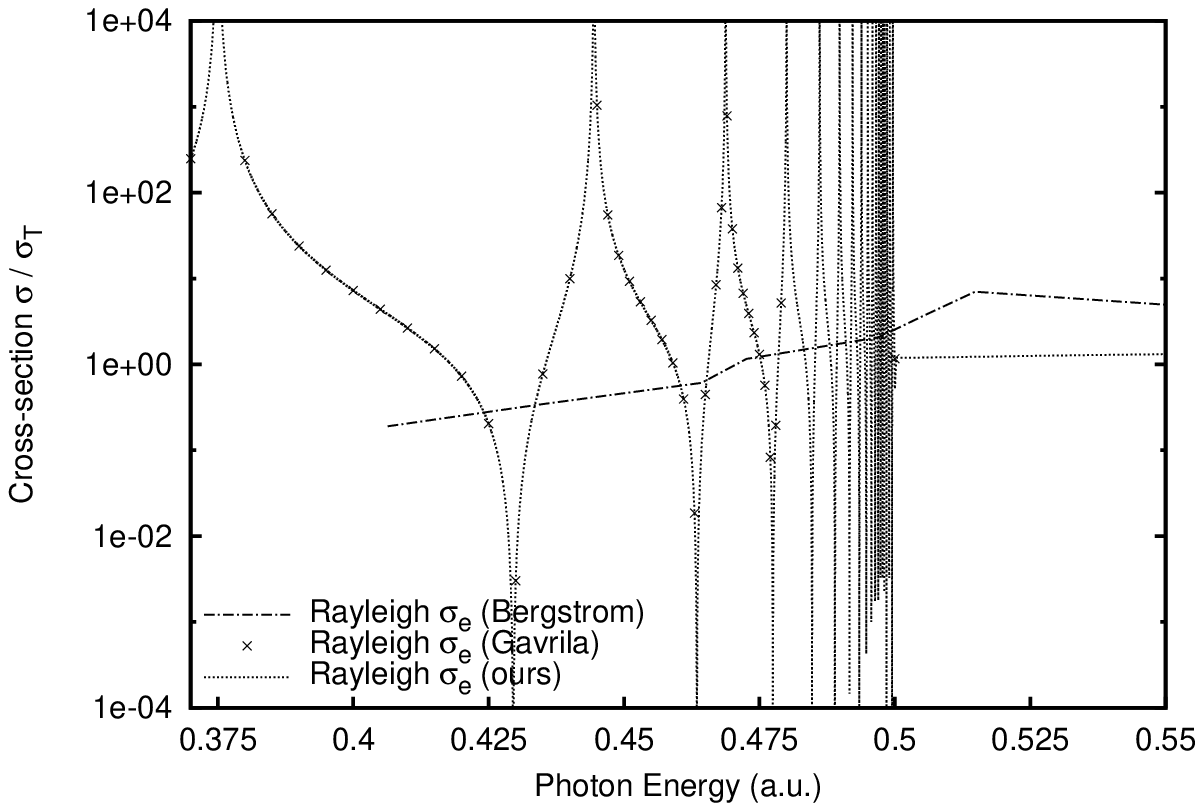}
\label{fig:rayleighzoom}
\end{figure}

\begin{figure}[tbh]
\caption{Raman scattering cross-sections for
photon-hydrogen scattering (in units of $\sigma_T$).
These are shown as a function of incident photon energy $\omega$ in
atomic units.
Our results are shown and compared against the (below threshold only)
results for H($1s$) $\to$ H($2s$) of Sadeghpour and Dalgarno~\cite{sadeghpour92b}.
The cross-section for H($1s$) $\to$ H($3d$) is also shown.}
\includegraphics[width=85mm]{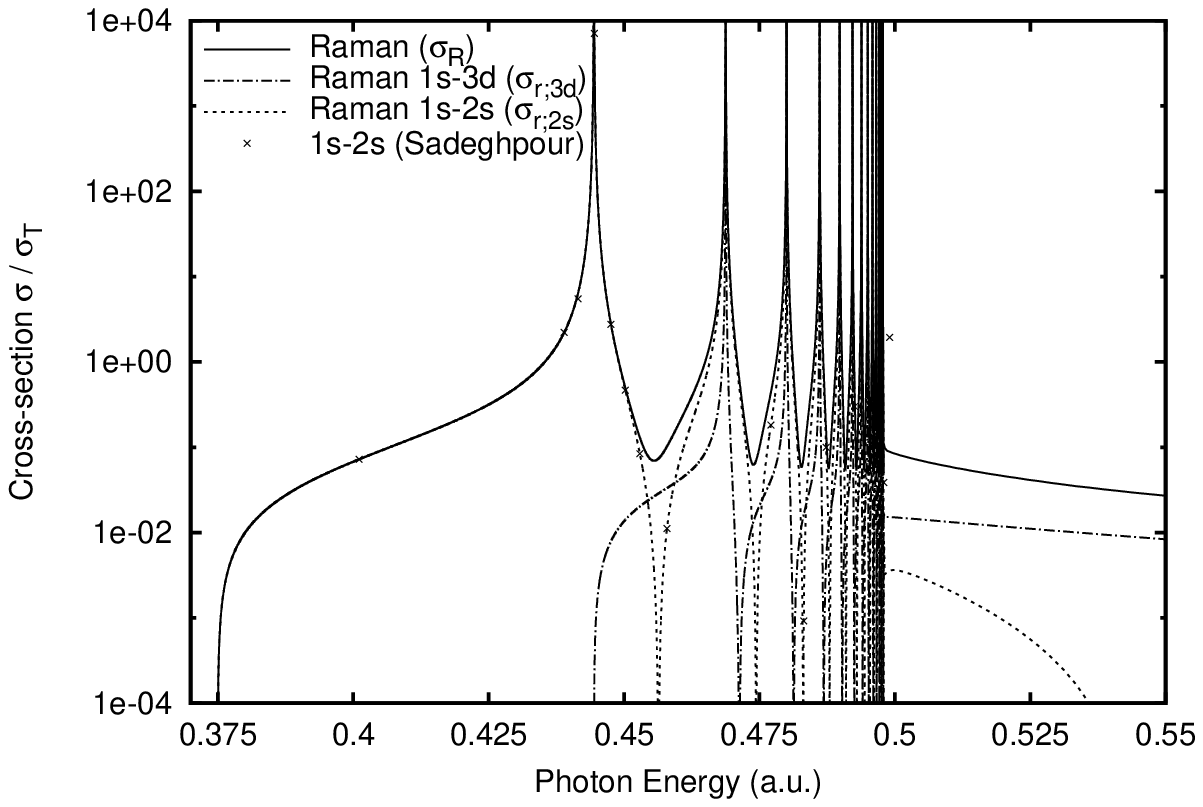}
\label{fig:ramanzoom}
\end{figure}

Fig.~\ref{fig:rayleighzoom} shows the Rayleigh cross-section between frequencies $0.37-0.55$~a.u.,
which spans the atomic (physical) resonances and also up into the continuum.
Fig.~\ref{fig:ramanzoom} shows the Raman cross-section across the same frequency range. 
This gives a closer look at the behaviour of the Raman cross-section near threshold. 
Table~\ref{tab:crosssectiondata} gives three data points for Rayleigh and Raman cross-sections. The photoionization cross-section 
is also given for the frequency above threshold. Above threshold the cross-sections must be computed on a pseudostate energy. 
Therefore, the values at $0.52$ a.u. were determined by interpolating linearly between two pseudostate frequencies. 
This introduces an uncertainty in the value, resulting in only few significant figures. Table~\ref{tab:crosssectiondata} gives 
the cross-section at $\omega = 0.52$a.u. as calculated for three different basis sets to show the convergence. Both in Raman and Compton 
this shows that the convergence is not monotonic with respect to the number of Laguerres in the basis set, and will be examined in 
future work.
\begin{table}[tbh]
\caption{The Rayleigh, Raman, Photoionization and Compton cross-sections  (in units of $\sigma_T$) are given for three frequencies. 
The cross-sections at frequency $\omega = 0.52$ a.u. are presented for basis sets with $N=80$, $N=100$ and $N=120$ Laguerres
for each partial wave.}
 \begin{tabular}{cccccc} \hline \hline
  $\omega$ (a.u.) & $N$ & Rayleigh $\sigma_e/\sigma_T$ & Raman $\sigma_R/\sigma_T$   & Photoionisation $\sigma_I/\sigma_T$ & Compton $\sigma_C/\sigma_T$ \\ \hline
0.40 &  80 & 7.2448357561099925 & 0.0677208258644729 &    &  \\
0.40 & 100 & 7.2448357561099925 & 0.0677208258644729 &    &  \\
0.40 & 120 & 7.2448357561099925 & 0.0677208258644729 &    &  \\
0.45 &  80 & 12.769933387763700 & 0.5657804519972952 &    &  \\
0.45 & 100 & 12.769933387763700 & 0.5657804519972952 &    &  \\
0.45 & 120 & 12.769933387763700 & 0.5657804519972952 &    &  \\
0.52 &  80 & 1.244   & 0.0456   & $8.531\times 10^6$ & 0.1524 \\
0.52 & 100 & 1.242   & 0.0462   & $8.530\times 10^6$ & 0.1517 \\
0.52 & 120 & 1.240   & 0.0461   & $8.528\times 10^6$ & 0.1519 \\ \hline \hline 
 \end{tabular}
\label{tab:crosssectiondata}
\end{table}

\section{Convergence of cross-sections with Laguerre-basis size}

\begin{figure}[tbh]
\caption{Convergence of Rayleigh and photoionization scattering cross-sections
for photon-hydrogen scattering (in units of $\sigma_T$).
The $\sigma_{e}$ and $\sigma_{I}$ are shown as a function of
incident photon energy $\omega$ in atomic units
(ie. up to $\approx 2700$~eV) for various Laguerre-basis set sizes $N$.
Our results broadly agree with previous (purely analytic) results as shown. 
Two plots are given to show the behaviour of Rayleigh and photoionization (left) as well as 
to highlight the small peak in the Rayleigh cross-section (right) by plotting on a smaller linear scale.
}
\includegraphics[width=85mm]{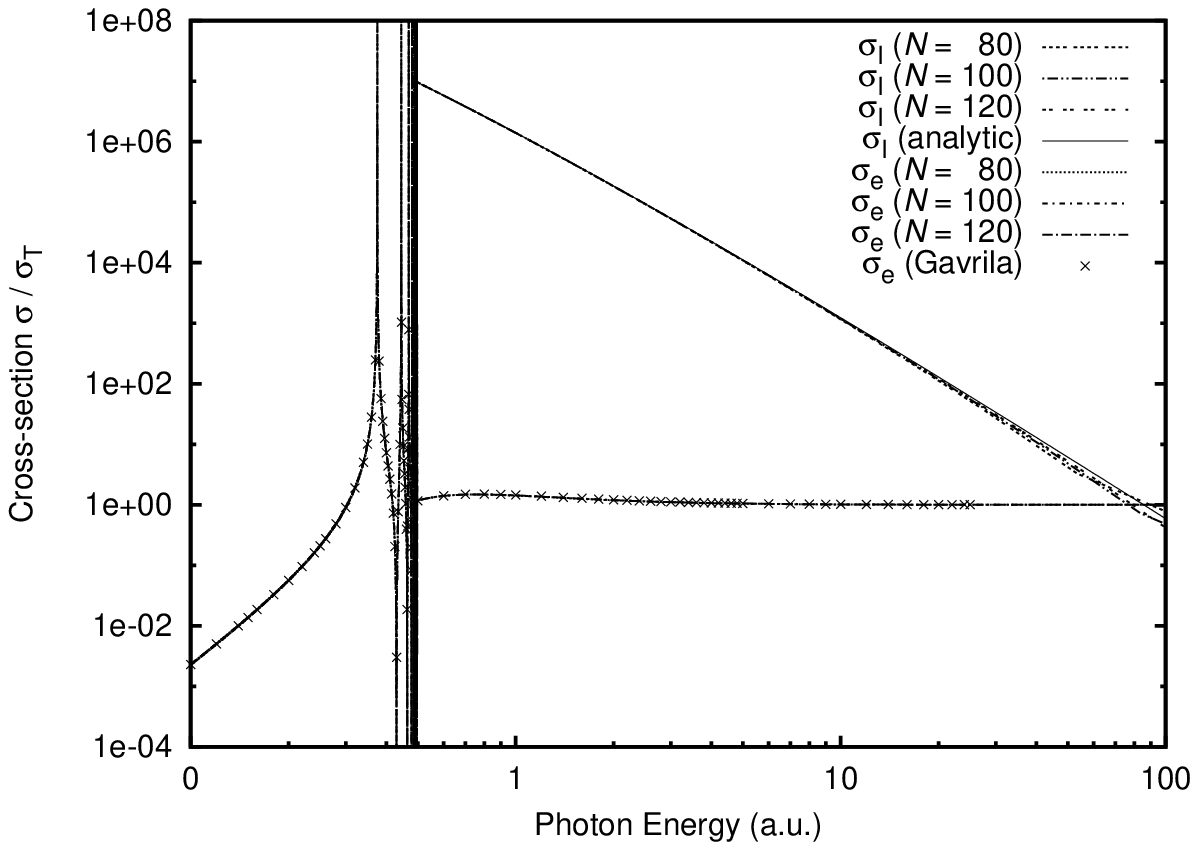}
\includegraphics[width=85mm]{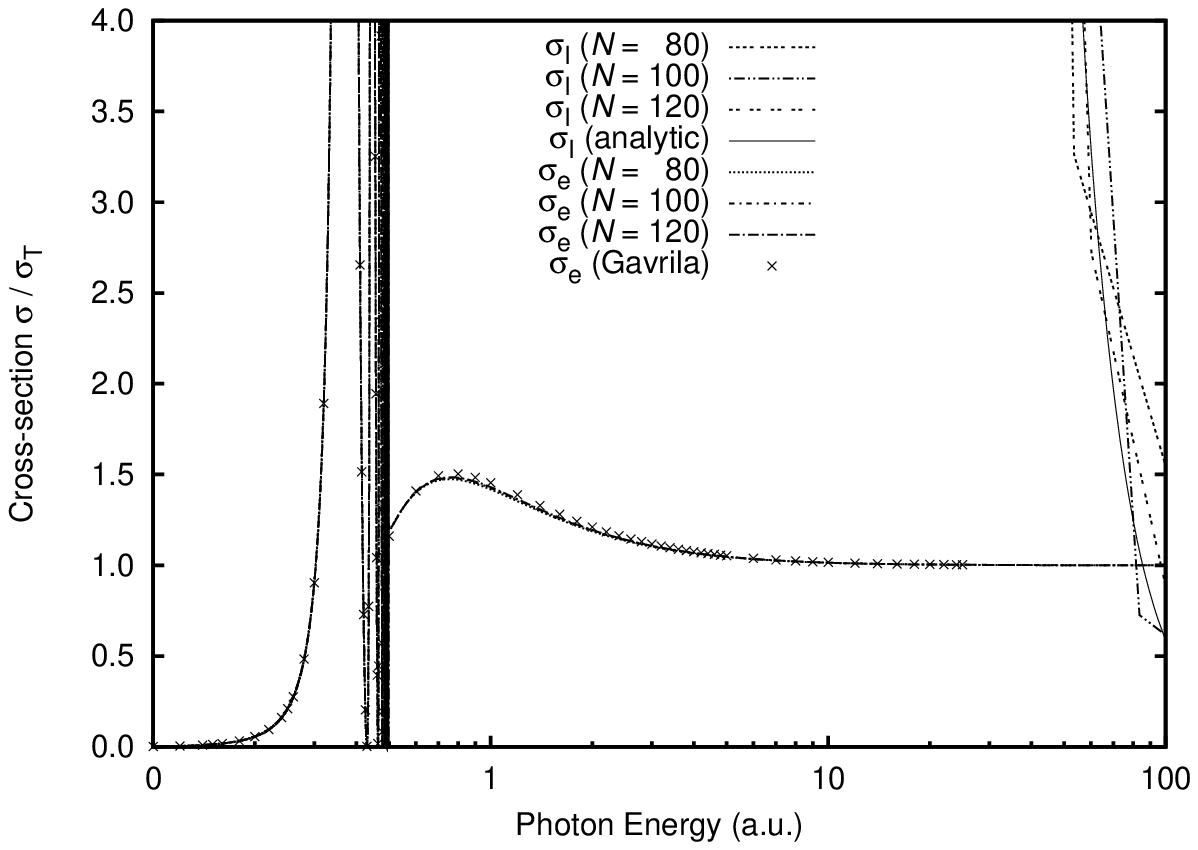}
\label{fig:rayleigh-converges}
\end{figure}

We ran off a series of calculations with the inclusion of various 
numbers of Laguerre functions, $N_\ell$ for each partial-wave.
The convergence of the Rayleigh and photoionization scattering
cross-sections are shown in Fig.~\ref{fig:rayleigh-converges}.
Note that the highest three eigenenergies in the $N=120$ calculation
are $E_{119p}=59.7669$, $E_{120p}=119.823$, $E_{121p}=357.323$~a.u.,
and thus it is no surprise that the polarizabilities develop
observable kinks around the $\omega \approx 100$~a.u. range.

\begin{figure}[tbh]
\caption{Convergence of Raman scattering cross-sections for
photon-hydrogen scattering (in units of $\sigma_T$) for
various Laguerre-basis set sizes $N$.
These are shown as a function of incident photon energy $\omega$ in
atomic units (ie. up to $\approx 2700$ eV).
Our results are shown for the sum of the individual cross-sections
$\sigma_{R}(\omega)$ (left) and for the $L_j=0$ or $L_j = 2$ contributions (right).
The total cross-section (left) has a bi-modal decay above
threshold due to the two different sets of contributions.
Note also that there are residual issues with the convergence
of the (relatively small) Raman cross-sections above threshold
due to the small pollution of the highest-lying `bound' state/s
which have a small mixture of pseudostate character which should
really be contributing to the (much larger) Compton cross-sections.
This can be seen by comparing the frequency location of the two Raman peaks
in the (right) plot against the location of the equivalent two Compton peaks in
Fig.~\ref{fig:comptoncompare}(right).
}
\includegraphics[width=85mm]{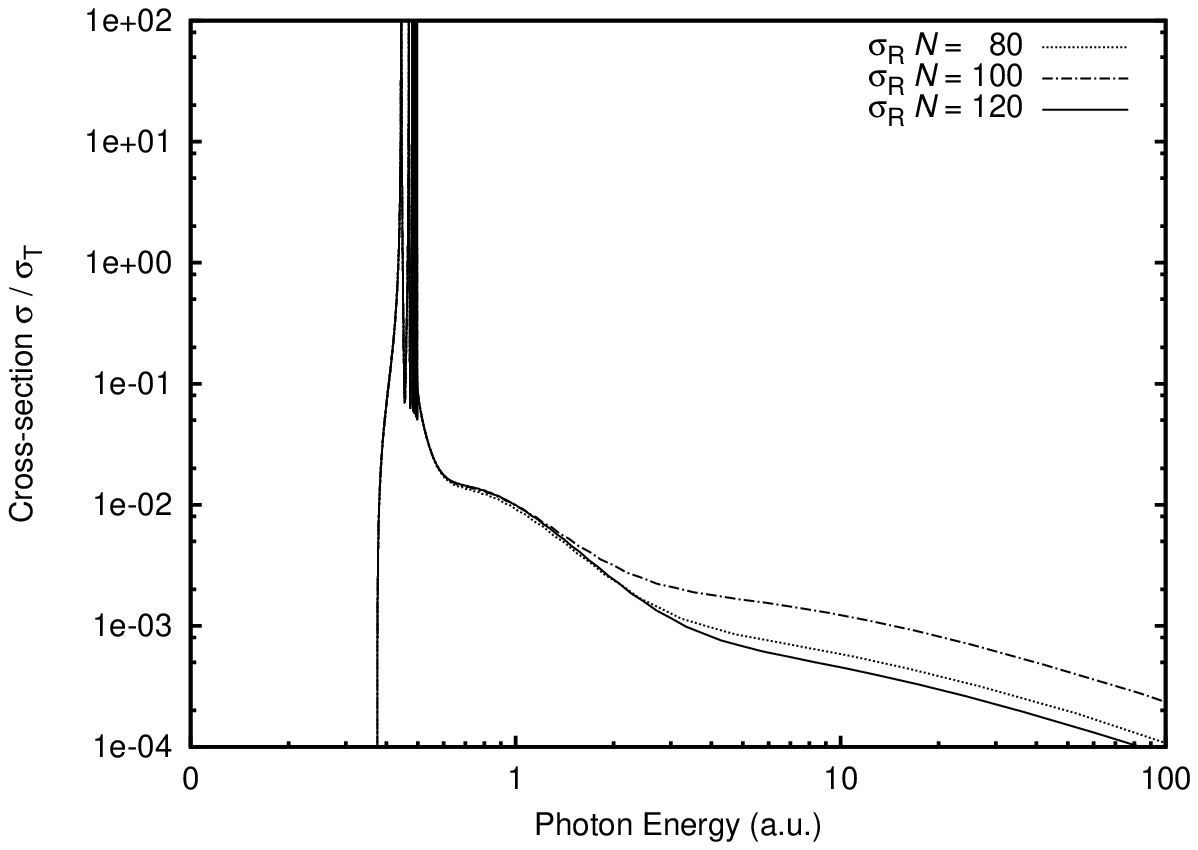}
\includegraphics[width=85mm]{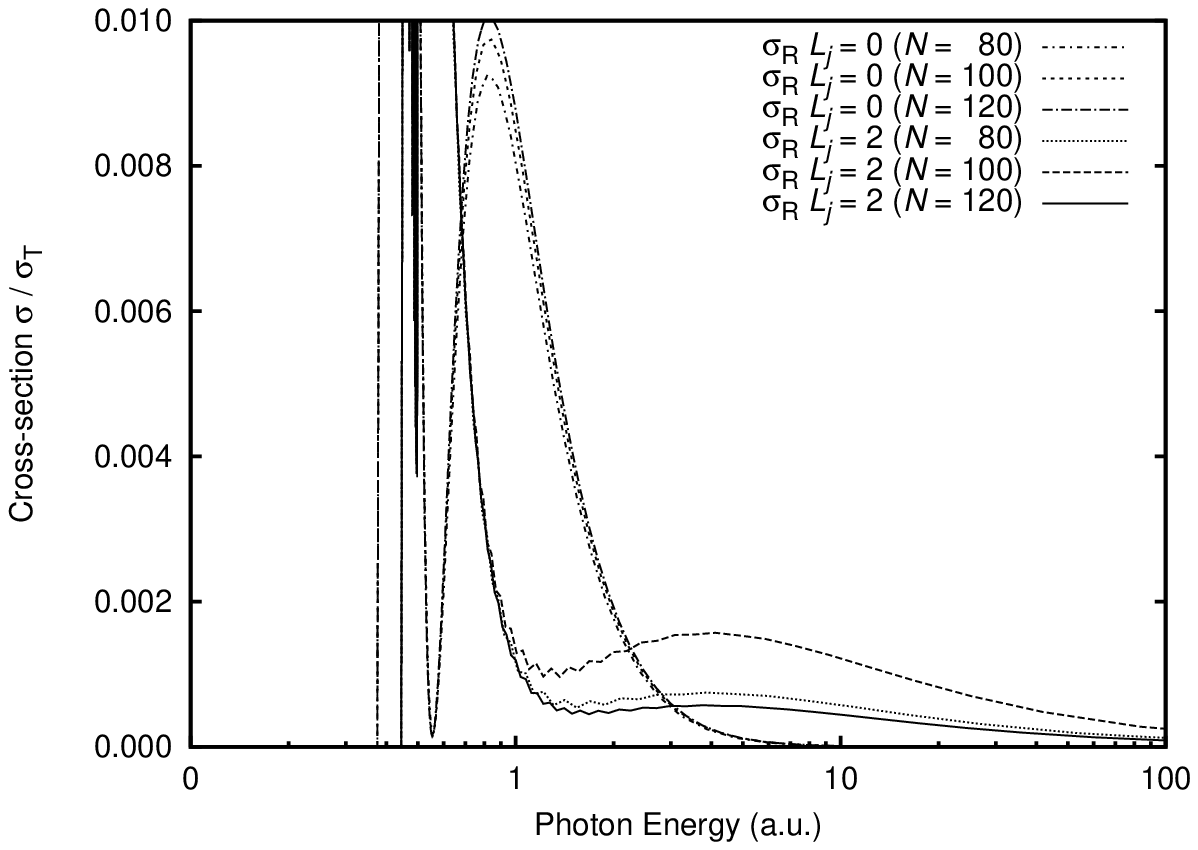}
\label{fig:ramanconvergence}
\end{figure}

The convergence of the Raman scattering cross-sections are shown
for the total 
$\sigma_{R}(\omega) = \sum_{(j;E_j < \Delta_N)} \sigma_{r;j}(\omega)$
for various $N_\ell$ (left) and the summed cross-section when 
only the $L_j = 0$ or $L_j = 2$ contributions are included (right).
Fig.~\ref{fig:ramanconvergence} shows that the $L_j = 0$ contribution 
rapidly declines just above threshold, and then increases rapidly.
The $L_j = 2$ contribution exhibits a small peak which changes with 
basis set size. This uncertainty in the (relatively small) cross-section 
will be studied further in future work.

\begin{figure}[tbh]
\caption{Convergence of Compton scattering cross-sections
for photon-hydrogen scattering (in units of $\sigma_T$)
for various Laguerre-basis set sizes $N$.
The summed cross-section is shown when only the
$L_j = 0$ contributions are included, which approximately
agrees with the results of Bergstrom \textit{et al.}~\cite{bergstrom93a} and
Drukarev \textit{et al.}~\cite{drukarev10a} which are also
shown. The summed cross-section when only the 
$L_j = 2$ contributions are included is also shown. 
The Compton cross-sections are plotted on a y-logscale (left) 
to show the behaviour of the $L = 2$ as well as the smaller $L = 0$ cross-sections.
This is also plotted on a linear scale (right), to show the 
convergence behaviour of the $N=120$, $N=100$ and $N=80$ calculations, 
not visible on a logscale.
}
\includegraphics[width=85mm]{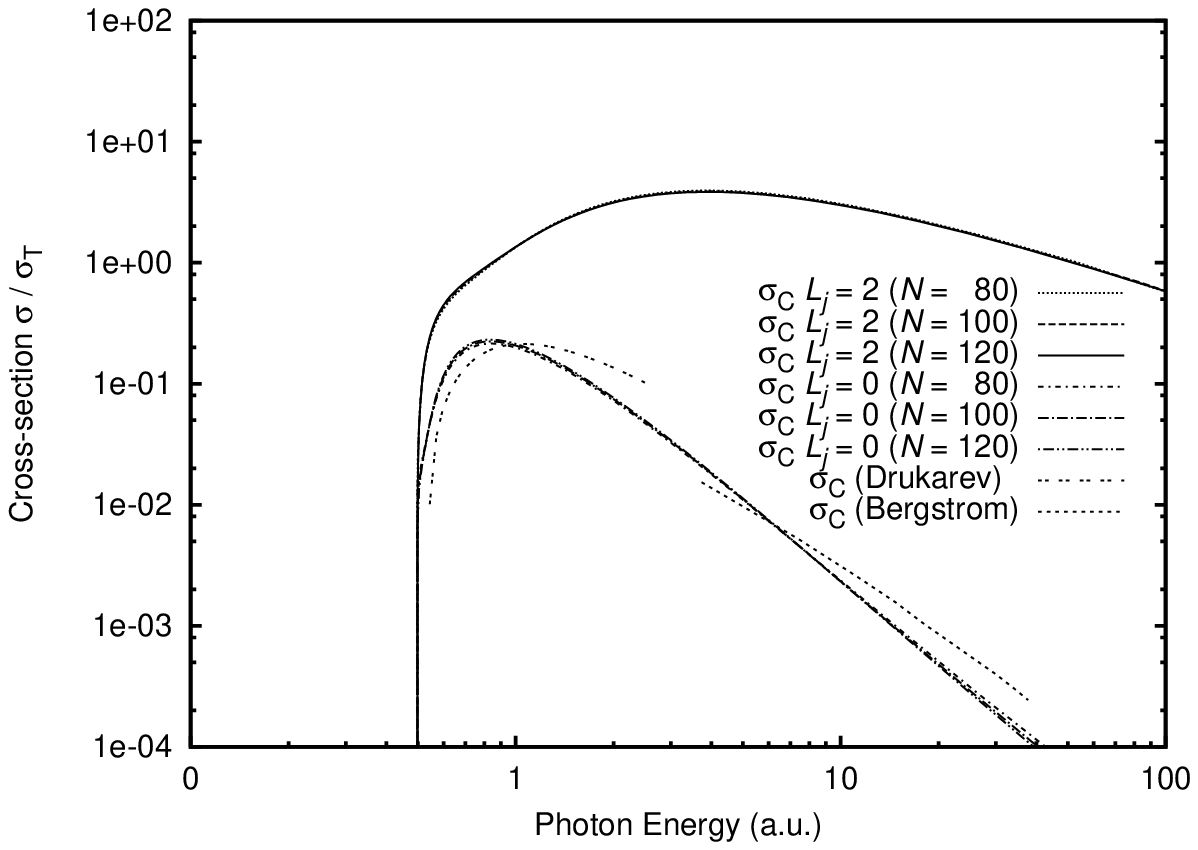}
\includegraphics[width=85mm]{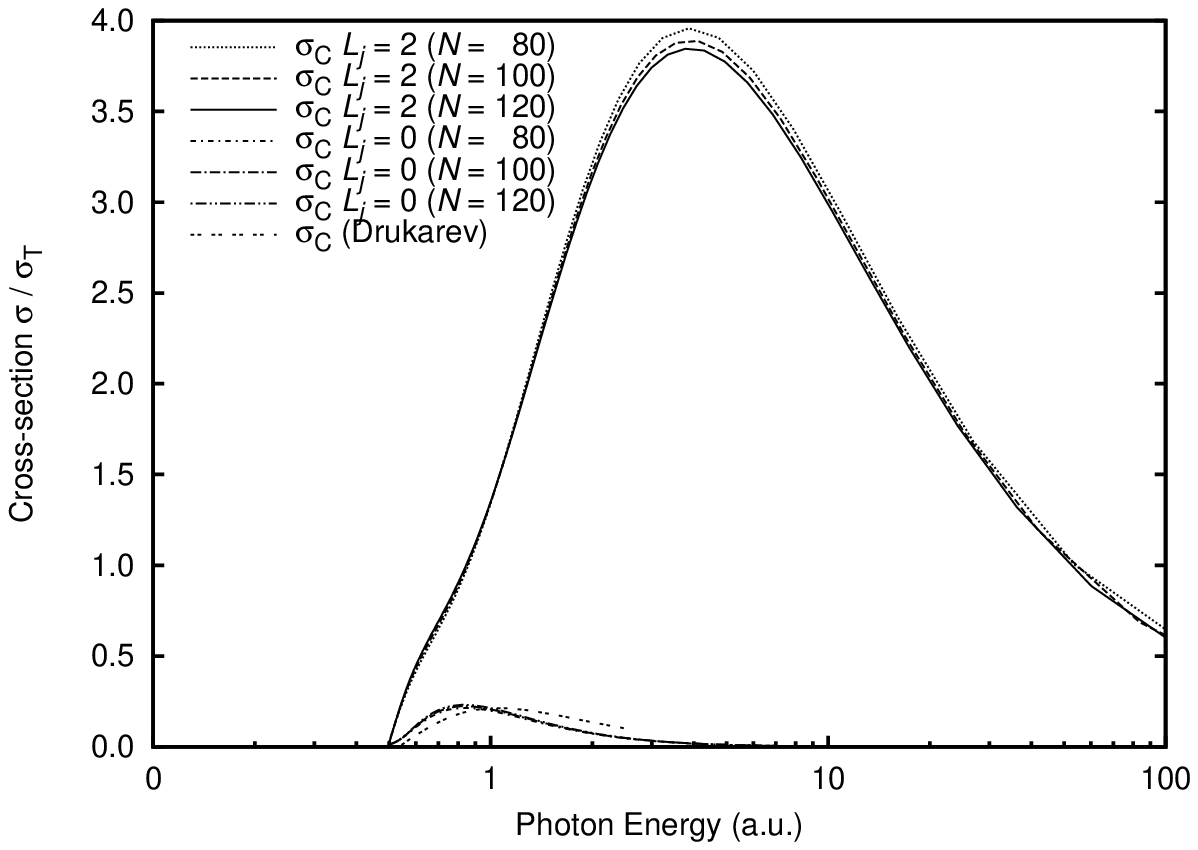}
\label{fig:comptoncompare}
\end{figure}

The convergence of the Compton scattering cross-sections are
shown (in Fig.~\ref{fig:comptoncompare}) when only the $L_j = 0$ or $L_j = 2$ contributions are included 
for various $N_\ell$. Here we show, as per paper, that the $L_j = 2$ contribution 
dominates the total Compton cross-section. Fig.~\ref{fig:comptoncompare} demonstrates that 
there is only a small change in the 
Compton cross-section when the size of the basis set is varied.

